# Integrating User's Domain Knowledge with Association Rule Mining

**Vikram Singh and Sapna Nagpal**
Department of Computer Science & Engineering
Chaudhary Devi Lal University, Sirsa (Haryana) INDIA.

**Abstract**

This paper presents a variation of Apriori algorithm that includes the role of domain expert to guide and speed up the overall knowledge discovery task. Usually, the user is interested in finding relationships between certain attributes instead of the whole dataset. Moreover, he can help the mining algorithm to select the target database which in turn takes less time to find the desired association rules. Variants of the standard Apriori and Interactive Apriori algorithms have been run on artificial datasets. The results show that incorporating user's preference in selection of target attribute helps to search the association rules efficiently both in terms of space and time.

***Keywords***: *Domain, association rule, data mining, Apriori, interactive Apriori.*

## 1. Introduction

Association rule is described as an associational relationship between a group of objects in a database [13]. Let D be a transaction database and I = {$i_1$, $i_2$, ..$i_m$} be an itemset. Transaction database D contains a sequence of transactions T = {$t_1$, $t_2$, .. $t_n$} (where T $\subseteq$ I) with a sole identifier. An association rule X→Y may be discovered in the data where X and Y are conjunctions of items and X ∩ Y = $\Phi$. The intuitive meaning of such a rule is that transactions in the database which contains the items in X tend to also contain the items in Y. The user supplies minimum support and confidence thresholds. The support of the rule X→Y represents the percentage of transactions from the original database that contain both X and Y. The confidence of the rule X→Y represents the percentage of transactions containing items in X that also contains items in Y. A rule that satisfies both minimum support and minimum confidence at the same time has been described as strong rule in the literature [2].

All the rules that meet the confidence threshold are reported as rules mined by the algorithm. The process of mining of association rules is broken up into two steps [3]:

(i) Find all the frequent itemsets in the database (i.e. the itemsets with support greater than the minimum support).

(ii) The confidence of the rule X→Y that satisfy minimum support is calculated as follows:

Confidence(X→Y) = support(XY)/support(Y)

### 1.1 Literature Survey

Association rules were first introduced by Agarwal et. al. in [1]. Their subsequent paper [3] discusses Apriori algorithm that is considered as one of the most important contributions to the subject of data mining. Although, other algorithms such as AIS [2] and SETM [7] are also available for mining association rules, yet Apriori remains the most widely used approach for generating frequent itemsets. The algorithm accomplishes the searching of frequent itemsets in recursive order. It first scans the database *D* and calculates the support of each single item in every record *I* in *D*, and denotes it as $C_1$. Out of the itemsets in $C_1$, the algorithm computes the set $L_1$ containing the frequent 1-itemsets. In the $k^{th}$ scan of the database, it generates all the new itemset candidates using the set $L_{k-1}$ of frequent (*k-1*) itemsets discovered in the previous scanning and denotes it as $C_k$. And the itemsets whose support is greater than the minimum support threshold are kept in $L_k$. This process is repeated until no new frequent itemsets are found.

Table 1: Dataset D

| $T_{id}$ | Items |
|---|---|
| 10 | AB |
| 20 | ABE |
| 30 | ABCE |
| 40 | CD |

The Apriori approach of searching frequent itemsets is explained with the database of Table 1. The algorithm assumes the minimum support threshold to be "*2*". Firstly, it initializes $C_1$ as the set of all items, takes count of elements in it, and puts in $L_1$ the elements satisfying the minimum support. Thereafter, set $C_2$ is generated using $L_1$ and count of the elements is computed from the scan of database *D*. The frequent itemsets from $C_2$ are kept in the set $L_2$. In the similar way, $L_3$ is generated. As there is a single itemset in $L_3$, the set $C_4$ is empty. So, this arithmetic comes to an end (min_support = 2). This has been explained in figure 1.

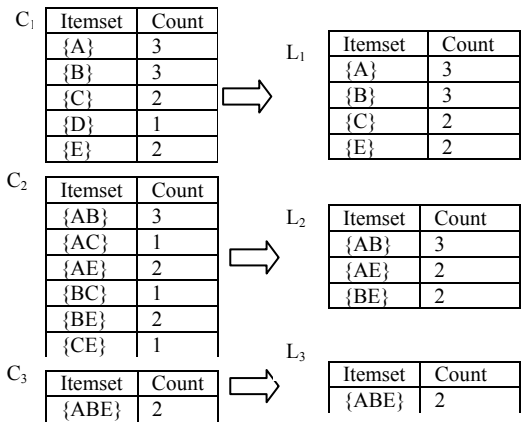

Fig. 1. Finding set of candidate and frequent itemsets with Apriori.

### 1.2 Generation of Association Rules

With the generation of frequent itemsets, the process of finding association rules begins. For every frequent itemset X, take non-empty subsets Y ($Y \subset X$, $Y \neq \Phi$) such that confidence ($Y \rightarrow (X-Y)) \geq$ minconf, an association rule "$Y \rightarrow (X-Y)$" is reported where confidence can be ascertained with Equation (1). support_count($Y \cup X-Y$) is the number of transactions containing itemsets ($Y \cup X-Y$) and support_count(Y) is the number of transactions containing itemset Y.

$$\text{confidence}(Y \rightarrow (X-Y)) = \frac{\sup port\_count(Y \cup (X-Y))}{\sup port\_count(Y)} \quad (1)$$

## 2. Mining with User's Guidance

Association rules are useful in data mining only if the mining analyst has a prior rough idea of what it is he is looking for. The key to knowledge discovery, therefore, is the user's domain knowledge. The domain expert has useful knowledge about the database which is not explicitly presented in the database [9]. This highlights the fact that there is no algorithm that will automatically furnish everything that is of interest in the database. An algorithm that finds a lot of useful rules will probably also find a lot of useless rules, while an algorithm that finds only a limited number of associations will probably also miss a lot of interesting information.

This discussion indicates that domain user must be involved in the process of finding association rules in the data sets. The domain user provides his suggestions and demands on the data mining result that tunes the process of rule discovery instead of proceeding in an unguided manner. A modification of Apriori that contains user's intervention in the processing of the algorithm is presented in the next section. The user at the first step provides some demands on the mining result that basically indicates what the user wants to see in the result. Accordingly the database is scrutinized on the particular attributes and it becomes the working database of the algorithm. Then the algorithm searches for associations among the attributes selected by the user. In this manner the user gets the association rules without exploring the whole database.

## 3. Interactive Association Rule Mining

The approach of user interactive association rule mining is embodied in IAR algorithm. The IAR is a variation of Apriori algorithm. The Apriori algorithm typically identifies the patterns that occur in the whole database. But what if the user is interested in particular attributes and wants to check if there is some associational relationship containing the attributes in the database. In such case it is irrelevant to do exhaustive search in the database. The IAR algorithm includes interaction points for the domain user to give attribute specification if any. The database is then scrutinized according to the specified attribute(s) i.e. the transactions not containing the attributes given by the user are excluded and a working database is created. With this subset of the dataset, the Apriori procedure searches for frequent large itemsets. Although the search dataset is scrutinized but the support for the potential large itemsets is calculated with respect to the original database. The Interactive Association Rule (IAR) algorithm is presented in Fig 2.

```
D′ := subset of D containing transactions having the
       attributes specified by the user.
       // (D′ is the working database)
L₁:= {frequent 1-itemsets};
k:=2;    //k represents the pass number.
while(L_{k-1} ≠ Φ )
     C_k := new candidates of size k
     generated from L_{k-1}
     for all transactions t ∈ D
     increment count of all candidates in C_k
     that are contained in t
     L_k :=  all candidates in C_k
     with minimum support
     k := k+1
Report U_k L_k as the discovered frequent itemsets
```

Fig. 2. The Interactive Apriori Algorithm (IAR).

Since searching the database for associational relationship is heavy task in large datasets, the time is saved as irrelevant records (in which user is not interested) are excluded from the database. The attributes in the database are randomly distributed; it may reduce the size of working dataset from half to even more fraction.

Considering the same example database of Table 1, here is how IAR algorithm works. At first it takes attribute preferences from the user. Suppose the user is interested in attribute B and wants to see if there is any frequent itemset containing itemset B. The IAR algorithm, at the first step, scrutinizes the database and creates a working database D′ from D. D′ contains transactions containing attribute B only (table 2). D′ contains 3 transactions, $T_{id}$ 40 is not included in D′ as this doesn't contain B. Size of the working database is thus reduced and it takes less time in all the scans of the database in the searching process of the algorithm. As shown in figure 3, the size of $C_k$ and $L_k$ get reduced starting from the first step. In this way there is no need to do an exhaustive search of the database if the user is interested in knowing the associational relationship containing a particular attribute.

Table 2: Dataset D′

| $T_{id}$ | Items |
|---|---|
| 10 | AB |
| 20 | ABE |
| 30 | ABCE |

### 3.1 Generation of Association Rules using frequent itemsets

The frequent itemsets found in the previous step are used to generate association rules. All the permutations and combinations of the items present in the frequent itemsets are considered as candidates for strong rules. A lot of rules will be generated in this way. A strong rule is one which has atleast minimum confidence which is computed by the Eq. (1) (see section 1.1).

It is important to note that the discovered rules contain the user specified attributes on the LHS and derives other attributes in the database. If such a rule possesses high confidence level then it could be valuable in the marketing context for the organization. In this way a lot of time can be saved and the user trusts more in the discovered rules.

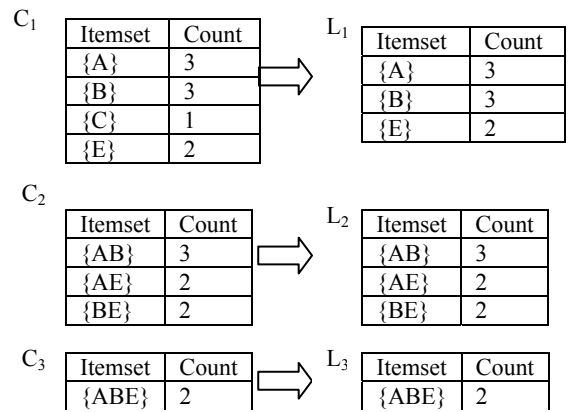

Fig. 3. Finding set of candidate and frequent itemsets with IAR.

## 4. Experimentation

For the purpose of performance evaluation IAR algorithm in discovering frequent itemsets, both Apriori and IAR have been run on the same platform under same conditions. Various parameters were computed for the purpose of comparison and the results have been shown in tables 3 and 4 and figure 4. The experimental runs have been conducted with two support levels and different sized datasets. It has been found that the IAR algorithm always takes less time and storage space than the standard Apriori. The interesting information can be mined in a shorter time. The test dataset has 7 attributes. The data was generated by artificial transactions to evaluate the performance of the algorithm over a range of data characteristics. The attributes are numbered starting from 1 and going in sequence. Any database of real world can be used with this algorithm by converting the attribute names to 1, 2, 3 and so on.

The algorithms use T-tree data structure to store frequent item set information. A T-tree is a "reverse" set enumeration tree where each level of the tree is defined in terms of an array. The storage requirement for each node (representing a frequent item set) in the T-tree is 12 bytes for a) reference to T-tree node structure (4 Bytes), b) support count field in T-tree node structure (4 Bytes) and c) reference to child array field in T-tree node structure (4 Bytes) [8].

Both the algorithms were compared with respect to the number of nodes in the T-tree structure, updates required to in T-tree to find large itemsets and the storage of T-tree in bytes as shown in Table 3 and 4. However the most important factor is time. IAR always takes less time than Apriori. The time comparison of both the algorithms is shown in Figure 4. It must be noted that the time taken and other parameters may differ for different runs as the data is generated randomly. Also the behaviour of IAR need not be same for different attributes specified by the user. But it always takes less time and storage than Apriori. It must also be noted that IAR does not do exhaustive search instead it finds association rule containing the attributes specification given by the user.

Table 3: Values of parameters with support level 20 percent

| Data Size | Number of frequent itemsets | | Number of nodes in T-tree | | Number of Updates required in T-tree | | Storage of T-tree in bytes | |
|---|---|---|---|---|---|---|---|---|
| | Apriori | IAR | Apriori | IAR | Apriori | IAR | Apriori | IAR |
| 2K | 31 | 15 | 43 | 20 | 26458 | 11722 | 496 | 244 |
| 10K | 32 | 13 | 45 | 19 | 132503 | 48566 | 504 | 212 |
| 30K | 28 | 13 | 41 | 19 | 336547 | 128218 | 476 | 248 |
| 50K | 30 | 15 | 42 | 18 | 574843 | 240544 | 484 | 272 |
| 120K | 28 | 15 | 41 | 21 | 1346085 | 589970 | 476 | 280 |

Table 4: Values of parameters with support level 30 percent

| Data Size | Number of frequent itemsets | | Number of nodes in T-tree | | Number of Updates required in T-tree | | Storage of T-tree in bytes | |
|---|---|---|---|---|---|---|---|---|
| | Apriori | IAR | Apriori | IAR | Apriori | IAR | Apriori | IAR |
| 2K | 20 | 9 | 33 | 14 | 22630 | 8883 | 324 | 148 |
| 10K | 17 | 7 | 33 | 13 | 110607 | 40556 | 276 | 144 |
| 30K | 15 | 5 | 30 | 10 | 291806 | 85866 | 240 | 140 |
| 50K | 15 | 5 | 30 | 10 | 482533 | 141980 | 240 | 140 |
| 120K | 15 | 5 | 30 | 10 | 1167228 | 342575 | 240 | 140 |

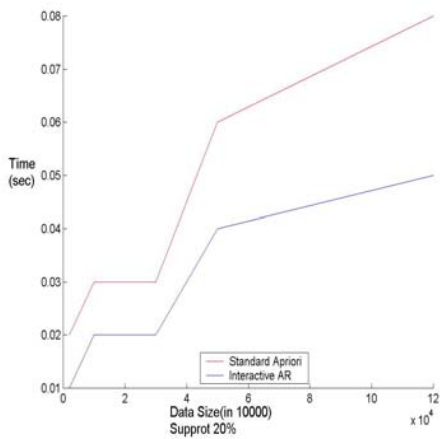

(a)    Support level 20%

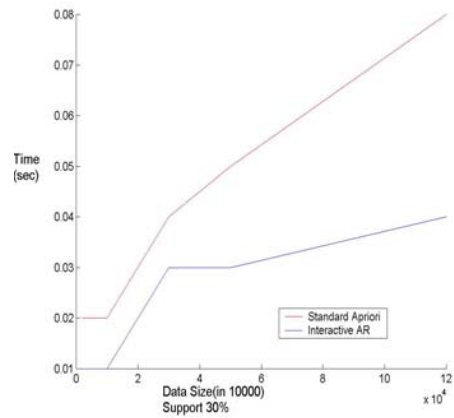

(b)    Support level 30%

Fig. 4. Temporal performance of Apriori (red - upper) and IAR (blue – lower) ((a) & (b)).

## 5. Conclusion

Among the various data mining techniques, rules are the most appropriate for integrating human opinions, because human thoughts can be converted into rules relatively easier than into some other form. User's suggestions and demands can be incorporated in the process to transfer domain knowledge that results in less and shorter iterations within the knowledge discovery loop.

This paper presents IAR algorithm that is a variation of standard Apriori algorithm to include user's role in finding interesting association among items in a database. The two algorithm are compared using different data sizes and support levels. The IAR always outperforms Apriori and the performance enhances as the data size increases. The domain user's knowledge may contribute the discovery of interested patterns.

**References**

[1] R. Agrawal, C. Faloutsos, and A. Swami, "Efficient similarity search in sequence databases", in thr Proceedings of the Fourth International Conference on Foundations of Data Organization and Algorithms, 1993, Vol. 730, pp. 69-84.

[2] R. Agrawal, T. Imielinski, and A. Swami, "Mining association rules between sets of items in large databases" in the Proceedings of the 1993 ACM SIGMOD International Conference on Management of Data, 1993, pp. 207-216.

[3] R. Agrawal, and R. Srikant, "Fast Algorithms for Mining Association Rule", in the Proceedings of the 20th International Conference on Very Large Databases (VLDB), 1994, pp. 487 – 499.

[4] M. Ankerest, "Human Involvement and Interactivity of the Next generation's Data Mining Tools", in ACM SIGMOD Workshop on Research Issues in Data Mining and Knowledge Discovery, Santa Barbara, CA, 2001.

[5] U. Fayyad, G.P. Shapiro, and P. Smyth, "The KDD Process for Extracting Useful Knowledge from Volumes of Data", in *Communications of ACM*, 1996, Vol. 39, pp. 27-34.

[6] J. Han, and M. Kamber, Data Mining: Concepts and Techniques, Morgan Kaufmann, 2006.

[7] M.A.W. Houtsma, and A.N. Swami, "Set-Oriented Mining for Association Rules in Relational Databases", in the Proceedings of the Eleventh International Conference on Data Engineering, 1993, pp. 25-33.

[8] T. J Lehman and M. J. Carey, "A Study of Index Structures for Main Memory Database Management Systems", in the Proceedings of the 12th International Conference on Very Large Data Bases, 1986, pp. 294-303..

[9] M. M. Owrang, and F. H. Grupe, "Using domain knowledge to guide database knowledge discovery", in Expert System with Application, 1996, Vol.10, pp. 173-180.

[10] S. Y. Sung, Z. Li, C.L. Tan, and P.A. Ng, "Forecasting Association Rules Using Existing Data Sets", in IEEE Transactions On Knowledge And Data Engineering, 2003, Vol. 15, pp. 1448-1459.

[11] D.X. Wang, X.Z. Hu, , X.P. Liu, and H. Wang, "Association Rule Mining on Concept Lattice Using Domain Knowledge", in the Proceedings of Fourth International Conference on Machine Learning and Cybernetics, Guangzhou, 2005, Vol.4, pp. 2151-2154.

[12] R. Wille, Restructuring Lattice Theory: An Approach Based on Hierarchies on Concepts in Ordered Sets, Rival, Boston, 1982, pp. 445-470.

[13] S. Zhang, S. Liu, D. Wang, J. Ou, and G. Wang, "Knowledge Discovery of Improved Apriori-based High-Rise Structure Intelligent Form Selection", in the Proceedings of the Sixth International Conference on Intelligent Systems Design and Applications, 2006, Vol.1, pp. 535-539.